\documentclass{article}
\usepackage{smacsmc2013}
\usepackage{times}
\usepackage{ifpdf}
\usepackage[english]{babel}
\usepackage{cite}
\usepackage{amsmath}
\usepackage{color}

\def\papertitle{Is the jet-drive flute model able to produce modulated sounds like Flautas de Chinos ?}
\def\firstauthor{Soizic Terrien}
\def\secondauthor{Christophe Vergez}
\def\thirdauthor{Patricio de la Cuadra}
\def\fourthauthor{Beno\^it Fabre}

\newif\ifpdf
\ifx\pdfoutput\relax
\else
   \ifcase\pdfoutput
      \pdffalse
   \else
      \pdftrue
\fi

\ifpdf 
  \usepackage[pdftex,
    pdftitle={\papertitle},
    pdfauthor={\firstauthor, \secondauthor, \thirdauthor, \fourthauthor},
    bookmarksnumbered, 
    pdfstartview=XYZ 
   ]{hyperref}

  \usepackage[pdftex]{graphicx}
  \graphicspath{{./figures/}}
  \DeclareGraphicsExtensions{.pdf,.jpeg,.png}

  \usepackage[figure,table]{hypcap}

\else 
  \usepackage[dvips,
    bookmarksnumbered, 
    pdfstartview=XYZ 
  ]{hyperref}  

  \usepackage[dvips]{epsfig,graphicx}
  \graphicspath{{./figures/}}
  \DeclareGraphicsExtensions{.eps}

  \usepackage[figure,table]{hypcap}
\fi

\hypersetup{
    colorlinks,%
    citecolor=black,%
    filecolor=black,%
    linkcolor=black,%
    urlcolor=black
}

\title{\papertitle}

 \fourauthors
   {\hspace{-1cm}\firstauthor} {\hspace{-1cm}Laboratoire de M\'ecanique et \\ \hspace{-1cm} d'Acoustique - CNRS UPR 7051 \\ \hspace{-1cm} Aix-Marseille Universit\'e,\\ \hspace{-1cm} Marseille, France\\%
     {\tt \href{mailto:author1@smcnetwork.org}{\hspace{-1cm}terrien@lma.cnrs-mrs.fr}}}
   {\secondauthor} {Laboratoire de M\'ecanique et \\ d'Acoustique  - CNRS UPR 7051 \\Aix-Marseille Universit\'e,\\ Marseille, France\\%
     {\tt \href{mailto:author2@smcnetwork.org}{vergez@lma.cnrs-mrs.fr}}}
   {\thirdauthor} { CITA, Centro de Investigaci\'on\\ en Tecnolog\'ias de Audio, Pontifi-\\cia Universidad Cat\'olica de Chile,\\ Santiago, Chile \\ %
     {\tt \href{mailto:author3@smcnetwork.org}{pcuadra@uc.cl}}}
   {\fourthauthor} {LAM-UPMC Univ. Paris 6,\\ UMR 7190, Institut Jean\\ le Rond d'Alembert,\\ Paris, France\\ %
     {\tt \href{mailto:author3@smcnetwork.org}{benoit.fabre@upmc.fr}}}

\begin{document}
\capstartfalse
\maketitle
\capstarttrue
\begin{abstract}
\textit{Flautas de chinos} - prehispanic chilean flutes played during ritual celebrations in central Chile - are known to produce very particular beating sounds, the so-called \textit{sonido rajado}. Some previous works have focused on the spectral analysis of these sounds, and on the input impedance of the complex resonator. However, the beating sounds origin remains to be investigated. 

Throughout this paper, a comparison is provided between the characteristics of both the sound produced by \textit{flautas de chinos} and a synthesis sound obtained through time-domain simulation of the jet-drive model for flute-like instruments. Jet-drive model appears to be able to produce quasiperiodic sounds similar to \textit{sonido rajado}. Finally, the analysis of the system dynamics through numerical continuation methods allows to explore the production mechanism of these quasiperiodic regimes.

\end{abstract}
%

\section{Introduction}\label{sec:introduction}

\begin{center}
\begin{figure}[h!]
\includegraphics[trim=5cm 8cm 0cm 1cm, clip=true,width=\columnwidth]{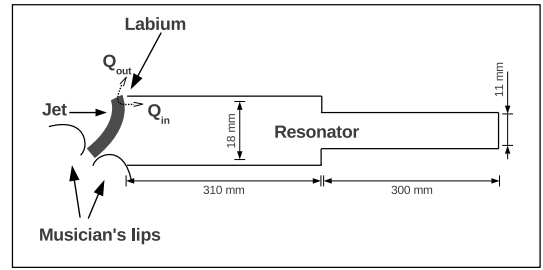}
\caption{Schematic representation of both the resonator of the studied \textit{flauta de chinos} and the excitation mechanism in flute-like instruments.}
\label{schema_excitateur}
\end{figure}
\end{center}

The \textit{Flautas de Chinos} are prehispanic chilean flutes still performed nowadays in ritual celebrations. The specificity of these intruments lies both in the production of very particular sounds (the so-called \textit{sonido rajado}, literally "torn sounds") and in the complex form of their resonator (highlighted in the schematic representation of the instrument provided in figure \ref{schema_excitateur}), formed by two pipes in series with different diameters.
Early works by Wright and Campbell \cite{Wright} have focused on the characteristics of the produced sounds (in terms of fundamental frequency, waveform and spectrum). Blanc \textit{et al.} \cite{Blanc_10} then attempted to describe and model the complex resonator.

However, the production mechanism of \textit{sonido rajado} remains unexplained. Moreover, to the authors knowledge, the behaviour of such instruments has never been investigated through the study of a physical model of sound production.

We do not develop in this paper an accurate, specific physical model of \textit{flautas de chinos}, but we propose to evaluate the ability of the classical jet-drive model for flute-like instruments to produce regimes similar to \textit{sonido rajado}. Moreover, such a study allows us to investigate the generation mechanism of those particular regimes.

The second section focuses on the description of the sound produced by \textit{flautas de chinos}. In section \ref{section_model}, we recall the state-of-the-art model for flute-like instruments, whose capacity to produce modulated sounds is evaluated in section \ref{section_time-domain simu} using time-domain simulation. Finally, we examine in section \ref{section_mechanism} the origin of such regimes.

\section{Analysis of \textit{Sonido Rajado}}
\label{section_son_reel}

The "\textit{sonido rajado}" produced by \textit{Flautas de Chinos} seems 
to be the combination of a very loud sound, a rich harmonic structure, and an outstanding beating component \cite{Wright,Blanc_10}. These beats can evoke the particular features of some quasiperiodic regimes, also produced for example in recorders or transverse flutes \cite{Coltman_06,Fletcher_76}. 

\begin{figure}[h!]
\includegraphics[width=\linewidth]{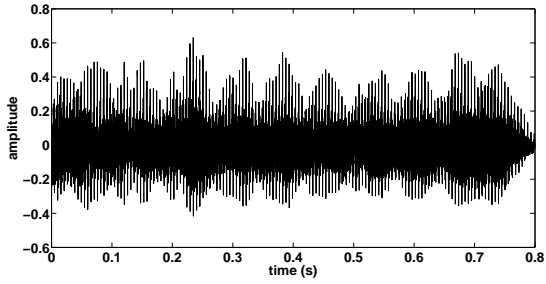}
\caption{Radiated sound produced by a \textit{flauta de chinos "puntera"}.}
\label{son_reel_vs_t}
\end{figure}

The sound produced by a musician playing a particular \textit{flauta de chinos} named \textit{puntera} -whose dimensions are specified in figure \ref{schema_excitateur}- has been recorded through a microphone located outside the instrument, and is represented in figure \ref{son_reel_vs_t}. One can note that this radiated sound shows strong envelope modulations, whose modulation frequency can be roughly approximate between 13 and 15 Hz.

To understand the production mechanism of these particular sounds, the knowledge of the regime nature is an essential information. However, the study of spectrum and waveform may be insufficient to distinguish, for example, a quasiperiodic regime from a chaotic regime. The computation of the Poincar\'e section (see for example \cite{Nayfeh,Gibiat}) is a commonly used method that allows to reveal the nature of the regime. However, in the present case, both the small length of the recorded signal and its nonstationarity prevent the use of this tool.

Return times can be computed to estimate periodicity at different time scales in embedded time series. The so-called return times are the times for which the time series returns to its location in the pseudo phase space \cite{Zou}. When this is done by chosing any point of the time series as a reference point, an histogram can be computed, which represents the occurrence of the different return times. Strong peaks in the histogram indicate periodicity.
According to Slater's theorem \cite{Zou}, for a noise-free signal, a quasiperiodic regime is characterized by the presence of three peaks in the histogram, corresponding to three "return times". Two of them are respectively related to the "base frequency" and to the "modulation frequency", whereas the third is the sum of the two others. In experimental time series, noise is inevitable. Its influence has been studied on return times by Zou \cite{Zou} who concludes that other peaks are likely to occur, but the sum relationship still holds.

\begin{figure}[h!]
\includegraphics[width=\linewidth]{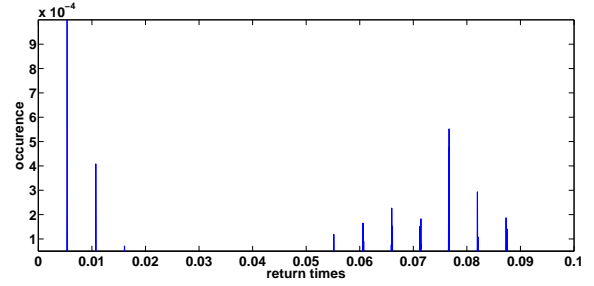}
\caption{Histogram representing the return times of the sound produced by the \textit{"puntera" flauta de chinos}.}
\label{son_reel_histogram}
\end{figure}

The histogram of the recorded sound, displayed in figure \ref{son_reel_histogram}, shows two principal peaks, respectively for the return times $t_1 = 0.0054$ s and $t_2 = 0.0767$ s. It highlights the presence, in the recorded sound, of two incommensurate frequencies: the "base frequency" $f_1 = 185$ Hz and the "modulation frequency" $f_2 = 13.04$ Hz.
All other return times of the histogram are either multiples of $t_1$ or combinations of the form $t_2 + p \cdot t_1$, with $p$ an integer. The histogram structure thus seems to satisfy Slater's theorem. Indeed, in the case of a chaotic regime, other peaks should appear in the histogram, related to return times which are neither multiples nor linear combinations of $t_1$ and $t_2$.

Therefore, the computation of return times allows us to conclude about the quasiperiodic nature of the \textit{sonido rajado} produced by \textit{flautas de chinos}.

\section{Model of flute-like instrument}\label{section_model}

\subsection{General mechanism of sound production}

Although instruments such as transverse flutes, flue organ pipes or the prehispanic chilean flutes studied here may appear different from each other, they involve the same general mechanism of sound production.
When the musician blows into the instrument, a naturally unstable jet is created at a channel exit. This channel is formed by the player's lips for transverse flutes and \textit{flautas de chinos}, whereas it is a part of the instrument for recorders and flue organ pipes (see figure \ref{schema_excitateur}).
This jet oscillates around an edge called "labium". The resulting interaction produces acoustic energy in the resonator, constituted by the air column contained in the instrument. In turn, the acoustic field thus created in the instrument perturbs the jet at the channel exit. This perturbation being convected and amplified from the channel exit toward the labium, it sustains the oscillation of the jet around the labium, and thus allows the emergence of self-sustained oscillations.

This functioning is detailed in \cite{Fabre_Kergo}, whose main elements are recalled in sections \ref{exciter} and \ref{resonator}. 

\subsection{Exciter}
\label{exciter}
\subsubsection{Hydrodynamics of the unstable jet}
Once a permanent operating regime of the instrument is reached, the perturbation of the jet (the so-called receptivity) is provided by the acoustic field, at the channel exit.
According to the empirical model proposed by de la Cuadra \cite{De_La_Cuadra_07}, the receptivity is represented as a transverse displacement $\eta(0,t)$ of the jet at the channel exit:

\begin{equation}
\eta(0,t) = \frac{h}{U_j} v_{ac} (t) ,
\label{eq:receptivity}
\end{equation}

\noindent
where $h$ is the channel height (\textit{i.e.} the opening height of the musician's lips), $v_{ac}$ is the acoustic velocity in the pipe, and $U_j$ is the jet central velocity (directly related to the fact that the musician blows hard or not).

As the jet is naturally unstable, this perturbation is amplified while convecting along the jet, from the channel exit to the labium. Based on the work of Rayleigh concerning instability of an infinite jet \cite{Rayleigh}, this phenomenon can be described in a simplified way as an exponential amplification of the perturbation with respect to the convection distance $x$:
\begin{equation}
\eta(x) = \eta(0) e^{\alpha_i x},
\label{eq:amplification}
\end{equation}

\noindent
Based on experimental works by de la Cuadra \cite{these_de_la_Cuadra}, an approximation of the amplification parameter $\alpha_i$ through the empirical expression: 
\begin{equation}
\alpha_i \approx 0.4/h,
\label{eq: coeff amplification}
\end{equation}
seems reasonable regarding the range of the Strouhal number $Str = \frac{fh}{U_j}$ considered throughout the paper.

Finally, we can express the jet transversal deflection at the labium:
\begin{equation}
\eta(W,t) = \eta(0,t-\tau) e^{\alpha_i W} = \frac{h}{U_j} v_{ac} (t-\tau)  e^{\alpha_i W},
\label{eq:deflection}
\end{equation}
where $W$ is the distance between the channel exit and the labium (corresponding to the length of the jet, highlighted in figure \ref{schema_excitateur}) and $\tau$ is the convection delay of the perturbation along the jet. Calling $c_p$ the convection velocity of this perturbation, $\tau$ is given by: $\tau = \frac{W}{c_p}$. Both theoretical and experimental works \cite{Rayleigh,Nolle_98,De_La_Cuadra_07} have shown that in the general case $0.3 U_j \leq c_p \leq 0.5 U_j $, which leads to:

\begin{equation}
\tau \approx \frac{W}{0.4U_j}
\label{eq:delay}
\end{equation}

\subsubsection{Nonlinear jet-labium interaction}
The interaction between the perturbed jet described above and the labium (see figure \ref{schema_excitateur}) makes the jet oscillate from one side to another of the labium. This oscillation causes a flow injection alternatively inside and outside the instrument. Based on the jet-drive model proposed by Coltman and followed by Verge \cite{Coltman_76,Verge_97}, these two (localised) flow injection in phase opposition are modelised as a dipolar pressure source:

\begin{equation}
\Delta p = -\frac{\rho \delta_d}{W H} \cdot \frac{d Q_{in}}{dt} ,
\label{eq:Delta_P}
\end{equation}

\noindent
In this equation, $\rho$ is the air density, $H$ the width of the jet and we note, as proposed by Verge \cite{Verge_94}, $\delta_d$ the effective distance between the two flow injection points. Based on both theoretical works and empirical estimation of parameters, Verge \cite{Verge_94} proposed the approximation $\delta_d \approx \frac{4}{\pi} \sqrt{(2hW)}$, which is adopted here.
$Q_{in}$ represents the flow injection in the pipe \cite{Fabre_Kergo}, given by:

\begin{equation}
Q_{in} = H  \int_{-\infty}^{y_{0}-\eta(t)} U(y) dy ,
\label{eq:Q_in}
\end{equation}

\noindent

In this expression, $y_0$ is the offset between the labium and the jet centerline, and $U(y)$ the velocity profile of the jet, representing by a Bickley profile:
\begin{equation}
U(y) = U_j sech^2(\frac{y}{b}) ,
\label{eq:Bickley}
\end{equation}

\noindent
where $b$ is the half width of the jet, related (under certain assumptions detailed for example in \cite{Segoufin_00}) to the height of the channel exit through $b=2h/5$.

By injecting equations (\ref{eq:Q_in}) and (\ref{eq:Bickley}) in equation (\ref{eq:Delta_P}), we finally obtain the expression of the pressure source that excites the resonator:

\begin{equation}
\Delta p_{src}(t) = \frac{\rho \delta_d b U_j}{W} \cdot \frac{d}{dt} \left[ tanh \left( \frac{\eta(t)-y_0}{b} \right) \right] .
\label{eq:source_aero}
\end{equation}

\subsubsection{Nonlinear losses}
Between the channel exit and the labium of flue instruments, the presence of an important transversal flow induced by the acoustic field in the pipe can cause, for high acoustic velocities, vortex shedding at the labium \cite{Fabre_96}. This phenomenon is modeled as nonlinear losses, and so represented by an additional nonlinear term $\Delta p_{los}$ in equation (\ref{eq:source_aero}):

\begin{equation}
\begin{split}
\Delta p(t)  =& \Delta p_{src} (t) + \Delta p_{los} (t) \\
 =&  \frac{\rho \delta_d b U_j}{W} \frac{d}{dt} \left[ tanh \left( \frac{\eta(t)-y_0}{b} \right) \right] \\
&- \frac{\rho}{2} \left( \frac{v_{ac}(t)}{\alpha_{vc}} \right)^2 sgn(v_{ac}(t))
\label{eq:source_aero_NL}
\end{split}
\end{equation}

\noindent
where $\alpha_{vc} \approx 0.6$ is a \textit{vena contracta} factor. 
\subsection{Resonator}
\label{resonator}
The acoustical response of the air column contained in the pipe, excited by the pressure source described above, is represented through the input admittance $Y_{in} = V_{ac}/\Delta P$ of this resonator. $V_{ac}$ and $\Delta P$ are respectively the frequency-domain expressions of the acoustic velocity at the pipe inlet and the pressure source.
The use of a modal decomposition of $Y_{in}$ is interesting as it allows an independant control of the different resonance mode characteristics. In the frequency domain, $Y_{in}$ is thus represented as a sum of resonance modes:
\begin{equation}
Y_{in} = \sum_m \frac{a_m j\omega}{\omega_m^2-\omega^2+j\omega \frac{\omega_m}{Q_m}} ,
\label{eq:decompo_modale}
\end{equation}
where $a_m$, $\omega_m$ and $Q_m$ are respectively the modal amplitude, the resonance pulsation and the quality factor of the $m^{th}$ resonance mode.

\begin{figure}
\centering
\includegraphics[width=\linewidth]{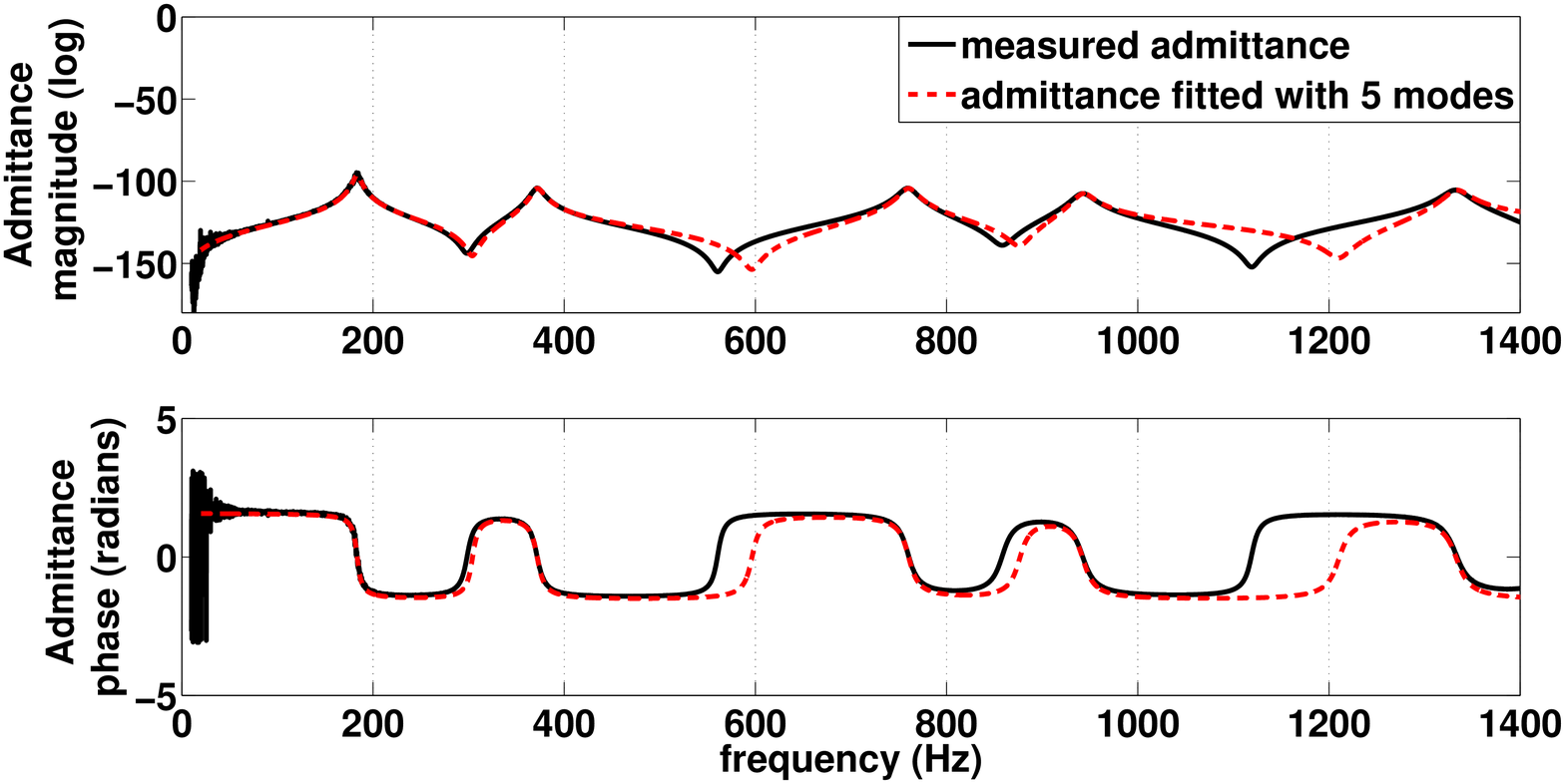}
\caption{Measured and fitted input admittance of a \textit{puntera} prehispanic flute. Only the five first resonance modes taken into account in the model are represented.}
\label{compa_admittance}
\end{figure}

As the particularity of the \textit{flautas de chinos} lies in the shape of their resonator, we use a modal decomposition of the measured admittance of the \textit{"puntera"} flute whose sound is studied in section \ref{section_son_reel}. The measure was realised thanks to an impedance sensor for wind instruments \cite{Macaluso_11}. The corresponding modal parameters (displayed in table \ref{table_coeff_modaux}) of the first five resonance modes are estimated by fitting the admittance calculated using equation (\ref{eq:decompo_modale}) to the measured one. 
One can note that the agreement between the original admittance and the fitted one, which are both displayed in figure \ref{compa_admittance}, can be improved  by increasing the number of resonance modes taken into account in the fit process. 
However, it considerabily increases the computation cost for the foregoing different resolution methods.

\begin{table}[t]
\begin{center}
\begin{tabular}{|c|c|c|c|}
\hline
mode number & $a_m$ & $\omega_m$ & $Q_m$ \\
\hline
1 & 11.39 & 1156.7 & 26\\
2 & 7.05 & 2342.8 & 34.4\\
3 & 9.55 & 4796.4 & 50.7\\
4 & 8.12 & 5943.4 & 52.9\\
5 & 12.93 & 8418.9 & 58\\
\hline
\end{tabular}
\end{center}
\caption{Fitted modal coefficients corresponding to the measured admittance of the \textit{puntera} flute.}
\label{table_coeff_modaux} 
\end{table}

\subsection{Numerical resolution of the model}
\label{method_resol}
The complexity of the model, due to its neutral nature and to the presence of nonlinear terms, imposes the use of numerical methods for solving the model. The neutral nature is related to the presence of a delayed derivative term, as discussed in \cite{CFA}.
In this study, we compare the results provided by two complementary methods: a classical time-domain simulation scheme, and a calculation method of periodic solutions coupled with a numerical continuation algorithm. Both are presented below and used in sections \ref{section_time-domain simu} and \ref{section_mechanism}.

\subsubsection{Time-domain simulations}
Time-domain simulations are performed in Matlab/Simulink, with a classical fourth-order Runge-Kutta method. Such an approach is interesting as it allows to access many kind of stable stationary regimes (for example static, periodic, quasiperiodic or chaotic) and to transients between these regimes.
However, results are very sensitive to initial conditions and to numerical parameters such as the sampling frequency. Moreover, unstable solutions cannot be computed: a slight disturbance (noise or even round-off errors, for example) is enough to make the system move away from these solutions. Therefore, it can be difficult to access, with such a method, to a global knowledge of the different oscillation regimes, essential to understand the system dynamics.

\subsubsection{Orthogonal collocation and numerical continuation}
Orthogonal collocation method allows to compute periodic solutions of dynamical systems. Its principle is based on the discretization of a single (unknown) period of the solution $x$ on $n$ intervals. On each interval, the solution is approximated by a polynomial of degree $d$. Projecting the model equations on this set of representation points leads to an algebraic system, whose unknowns are the value of $x$ at each discrete point, and the value of the period $T$.

This method, which allows to determine a solution $x_0$ for a set of model parameters $\lambda_0$, is coupled to a numerical continuation algorithm. Starting from $x_0$, it computes, using a Newton-Raphson prediction-correction method, the neighboring solution corresponding to a set of parameter $\lambda_0+\Delta \lambda$ \cite{Krauskopf}. Proceeding by successive iterations, one finally access to a complete \textit{branch} of periodic solutions. 

The computation of the Floquet multipliers allows subsequently to determine the stability properties of each point of the branch \cite{Nayfeh}. Indeed, according to Floquet theory, a periodic solution (\textit{i.e.} a point of the branch) is stable until all its Floquet multipliers lies in the unit circle. When a periodic solution loses its stability, the resulting regime observed after this bifurcation point depends on which way the Floquet multipliers leave the unit circle at the bifurcation point. For example, a quasiperiodic regime may appear if two complex conjugate Floquet multipliers leave the unit circle  \cite{Nayfeh}.

Giving access to the bifurcation diagram, which represents in an ideal case all the periodic and static solutions of the studied system, this method provides a more global knowledge of the system dynamics (coexistence of multiple solutions, unstable solutions ...), and thus permits an easier interpretation of phenomenon such as hysteresis or emergence of non periodic regimes (see for example \cite{CFA}).

The neutral nature of the model prevents the use of classical numerical continuation software (such as, for example, AUTO \cite{Doedel_81} or Manlab \cite{Cochelin_09}). We use here DDE-Biftool \cite{manuel_Biftool}, a software specifically developped for numerical continuation in delay differential equations, and its extension for neutral systems \cite{Barton_06}.

\section{Time-domain simulation: characteristics of the synthesis sound}
\label{section_time-domain simu}
Although we study here a very generic model of flute-like instruments, time-domain simulations lead to non periodic regimes which can perceptually recall the \textit{sonido rajado} produced by \textit{flautas de chinos}. 
Because of the few studies about parameter estimation in \textit{flautas de chinos}, some of the model parameters are chosen arbitrarily close to those currently uses in the case of a  model of an alto recorder \cite{CFA}, except for the excitation window area $WH$ and the labium offset $y_0$ which are here slightly higher than in the case of an alto recorder. As these parameters are fixed by the player in the case of \textit{flautas de chinos}, the chosen values (W = 1cm, H = 6mm and $y_0$ = 0.2 mm) seem reasonable. Parameter values used throughout the paper are provided in table \ref{table_parametres}.

\begin{table}[t]
\begin{center}
\begin{tabular}{|c|c|}
\hline
Parameter & Numerical value (S.I.)\\
\hline
$\alpha_i$ & 0.4/h = 400\\
\hline
$\delta_d$ & $\frac{4}{\pi} \sqrt{(2hW)}$ = 0.0057\\
\hline
b & 2h/5 = 0.0004\\
\hline
$y_0$ & 0.0002\\
\hline
h & 0.001\\
\hline
W & 0.01\\
\hline
$\rho$ & 1.2\\
\hline
$c_p$ & $0.4 U_j$\\
\hline
$\alpha_{vc}$ & 0.6\\
\hline
\end{tabular}
\end{center}
\caption{Parameter values used for numerical resolutions of the model.}
\label{table_parametres} 
\end{table}

A first simulation is achieved using a constant jet velocity $U_j = 39$ m/s, which would correspond to a case where the musician blows in a steady way. This value, which would be equivalent, using the stationary Bernoulli equation, to a mouth pressure of about 915 Pa, is of the same order of magnitude as the values measured in the player's mouth for the \textit{puntera} flute (between 886 Pa and 1546 Pa).

\begin{figure}[h!]
\includegraphics[width=\linewidth]{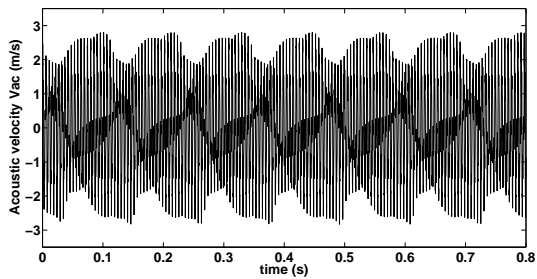}
\caption{Acoustic velocity computed through a time-domain simulation of the model for a constant jet velocity $U_j = 39$ m/s, represented as a function of the time. The sampling frequency used for computation is 22x44.1 kHz.}
\label{forme_onde_simu_Uj39}
\end{figure}

The computed acoustic velocity at the pipe inlet $v_{ac}$, represented in figure \ref{forme_onde_simu_Uj39} as a function of the time, shows, as the real sound (see figure \ref{son_reel_vs_t}), envelope modulations. A Fourier analysis of the signal envelop provides the modulation frequency $f_{mod}$ = 8.6 Hz. One can note that this value is of the same order of magnitude as the modulation frequency of the real sound, $f_2 = 13.04$ Hz (see section \ref{section_son_reel}), and coherent with the observations of Wright and Campbell \cite{Wright}. In the case of time-domain simulation, this frequency strongly depends on the the jet velocity $U_j$, and a slight shift of this parameter is enough to achieve the experimental value of $13.04$ Hz. However, the fact that the modulation frequency is related to the jet velocity is a significant difference compared to the real instruments behaviour. Indeed, as observed by Wright and Campbell \cite{Wright}, the modulation frequency shows only little variations with the pressure in the musician's mouth.

\begin{figure}[h!]
\includegraphics[width=\linewidth]{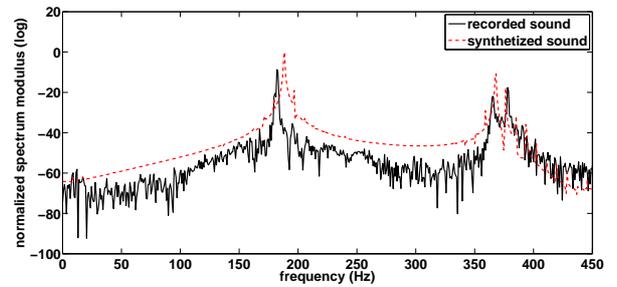}
\caption{Normalized spectrums of both the sound produced by a \textit{"puntera" flauta de chinos} (represented in figure \ref{son_reel_vs_t}) and the synthesis sound represented in figure \ref{forme_onde_simu_Uj39}.}
\label{compa_spectres}
\end{figure}

Although the physical model is not specific to the instrument studied here, and that some parameters can only be roughly estimated, the implementation of a measured admittance allows us to compare qualitatively, in figure \ref{compa_spectres}, the spectrum of the synthesis sound with that of the real sound. Such a comparison shows that both the fundamental and modulation frequencies (particularly visible around 370 Hz) of the synthetized sound are of the same order of magnitude as those of the real sound. One can note that the amplitudes can not be compared: indeed, we consider the spectrum of the sound radiated by the real instrument whereas we consider, for the model, the internal acoustic field (which would correspond to a measure under the labium of the instrument). Moreover, to make easier the qualitative comparison of the frequencies, the two spectra have been normalized with respect to their own maximum.

\begin{figure}
\includegraphics[width=\linewidth]{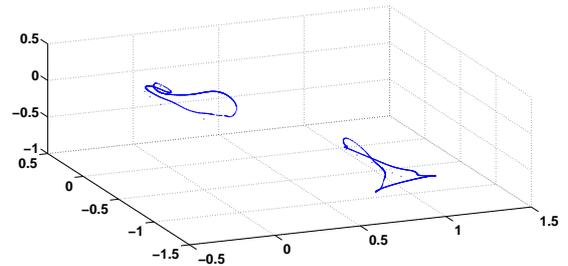}
\caption{Three-dimensional Poincar\'e section of the syntesis sound represented in figure \ref{forme_onde_simu_Uj39}. The closed curve is characteristic of a two-period quasiperiodic regime.}
\label{poincare}
\end{figure}

As for the real sound, one can wonder about the quasiperiodic or chaotic nature of the synthesis sound. The study of the Poincar\'e section allows us to distinguish these different kinds of regime from each other.
As shown on figure \ref{poincare}, the three-dimensional Poincar\'e section of the synthesis signal reveals a densely packed closed curve, which is characteristics of a two-period quasiperiodic solution (see for example \cite{Nayfeh}), and thus allows to conclude about the quasiperiodic nature of the simulated signal.

As a first conclusion, one can note that this study highlights the ability of the state-of-the-art physical model of flute-like instruments to produce regimes where some of their features such as quasiperiodic nature, fundamental and modulation frequencies, are similar to those of the sound produced by a \textit{flauta de chinos}.

\section{Quasiperiodic regime: generation mechanism}
\label{section_mechanism}
As it provides a more global knowledge of the system dynamics, the study of the bifurcation diagram of the model studied, which ideally represents all its stable and unstable periodic solutions as a function of one of its parameters, has recently shown its interest in the understanding of the functioning of musical instruments \cite{Karkar_11}.

\begin{figure}[h!]
\includegraphics[width=\linewidth]{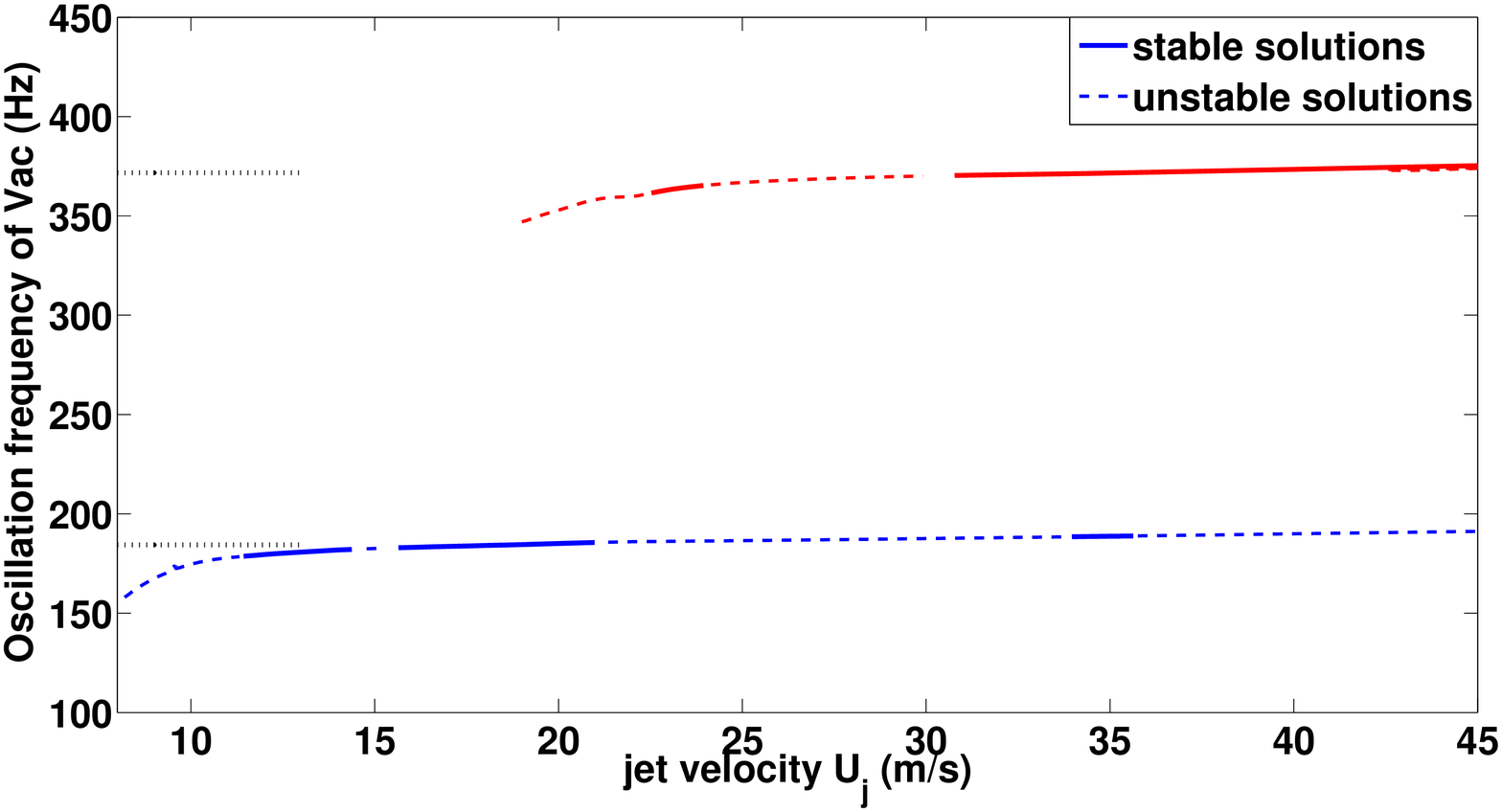}
\caption{Bifurcation diagram of the studied system, with two periodic solution branches corresponding to the first and the second registers. Abscissa: jet velocity $U_j$ (m/s). Ordinate: oscillation frequency (Hz). The horizontal dotted lines correspond to the two first resonance frequencies of the resonator.}
\label{diag_bif}
\end{figure}

The bifurcation diagram of the model displayed in figure \ref{diag_bif} represents the oscillation frequency of the different periodic solutions as a function of the jet velocity $U_j$. Solutions are computed for a range of the jet velocity $U_j$ for which quasiperiodic regimes may occur in time-domain simulations (see section \ref{section_time-domain simu}).
It shows the existence, in this range of $U_j$, of two periodic solutions branches: the first corresponds to the first register of the instrument (that is to say, to oscillations emerging from an instability at a frequency close to the first resonance frequency of the resonator), and the second is related to the second register. Solid and dotted lines respectively represent stable and unstable parts of periodic solution branches. 

In the previous section, we highlighted that time-domain simulation of the model reveals the existence of a quasiperiodic regime for $U_j = 39 m/s$.
For this value, the bifurcation diagram predicts that the first register is unstable, whereas the second register is stable.

\begin{figure}[h!]
\includegraphics[width=\linewidth]{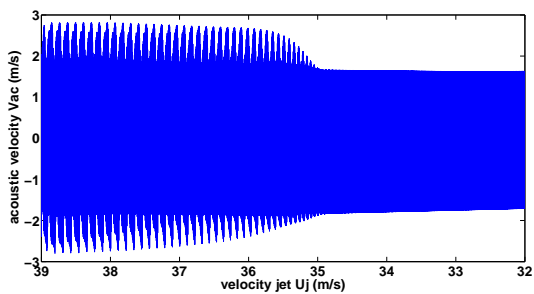}
\caption{Acoustic velocity computed through a time-domain simulation, for a linear decreasing ramp of the jet velocity $U_j$. The sampling frequency used for simulation is 22x44.1 kHz.}
\label{simu_rampe_Uj}
\end{figure}

In order to understand the origin of this quasiperiodic solution, we computed a second time-domain simulation, during which the jet velocity $U_j$ follows a decreasing ramp, from $U_j = 39 m/s$, to $U_j = 32 m/s$. The result, displayed in figure \ref{simu_rampe_Uj}, shows a gentle transition from the quasiperiodic regime studied in the previous section, to a periodic regime. A Fourier analysis of this periodic regime provides an oscillation frequency of 188.4 Hz, which is very close to the first resonance frequency $\frac{\omega_1}{2 \pi} = 184.4$ Hz, and thus allows to conclude that this periodic solution corresponds to the first register.

Thereby, the confrontation of figures \ref{diag_bif} and \ref{simu_rampe_Uj} suggests that the quasiperiodic solution is the result of the loss of stability of the first register (predicted by the bifurcation diagram at $U_j = 35.6 m/s$ - see figure \ref{diag_bif}) through a direct Neimark-Sacker bifurcation \cite{Nayfeh}, .

If the numerical tools used here do not allow computation and continuation of quasiperiodic solution branches, a further analysis of the stability properties of the periodic solutions permits to analyse the birth of quasiperiodic solutions (see section \ref{method_resol}). 

\begin{figure}[h!]
\includegraphics[width=\linewidth]{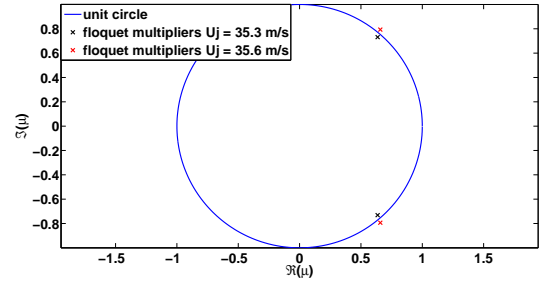}
\caption{Floquet multipliers represented in the complex plane, for two points of the first register periodic solution branch. The first point is located just before the loss of stability of the first register ($U_j = 35.3$ m/s) and the second one just after ($U_j = 35.6$ m/s).}
\label{Floquet_unit_circle}
\end{figure}

The Floquet multipliers associated to the first register, around the bifurcation point ($U_j = 35.6$ m/s) are shown in figure \ref{Floquet_unit_circle} for both the point of the branch corresponding to $U_j = 35.3$ m/s (just before the loss of stability) and the point corresponding to $U_j = 35.6$ m/s (just after the loss of stability). One can clearly observe that the loss of stability is the result of the crossing of the unit circle by two complex conjugate Floquet multipliers.

\begin{figure}[h!]
\includegraphics[width=\linewidth]{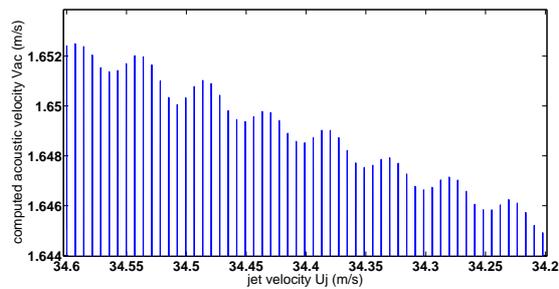}
\caption{Zoom of figure \ref{simu_rampe_Uj}: acoustic velocity computed through a time-domain simulation, for a linear decreasing ramp of the jet velocity $U_j$. The sampling frequency used for simulation is 22x44.1 kHz.}
\label{simu_rampe_Uj_ZOOM}
\end{figure}

This analysis not only confirms the generation of a quasiperiodic regime, but it also provides the value of the modulation frequency at the quasiperiodic regime threshold (\textit{i.e} at $U_j = 35.6$ m/s), through the computation of the Floquet exponents $\gamma_f$, which are related to the Floquet multipliers $\mu_f$ through \cite{Nayfeh}:

\begin{equation}
\gamma_f = \frac{1}{T} ln(\mu_f)
\label{exposants_Floquet}
\end{equation}

\noindent
where $T$ is the period of the periodic solution at the bifurcation point. 
According to Floquet theory, the imaginary part of the Floquet exponent corresponds to the modulation pulsation $\omega_{mod}$.
Here, such an analysis of the Floquet multipliers predicts a modulation frequency: $f_{mod}$ = $\omega_{mod}/2\pi$ = 26.4 Hz.
A further study of the synthesis signal envelop represented in figure \ref{simu_rampe_Uj} shows a threshold of the quasiperiodic regime at about $U_j = 34.2$ m/s (see a zoom of figure \ref{simu_rampe_Uj} in figure \ref{simu_rampe_Uj_ZOOM}). As in section \ref{section_time-domain simu}, a Fourier analysis of this envelop signal provides the modulation frequency related to the quasiperiodic regime: $f_{mod simu} = 26.3$ Hz. Although the quasiperiodic regime threshold seems to be underestimated in the time-domain simulation (which may be a consequence of both the sampling and the dynamics of the control parameter $U_j$ \cite{bergeot2012prediction}), the modulation frequency shows good agreement with those predicted by the Floquet analysis.

\section{Conclusion}

This study, rather than proposing a specific physical model of \textit{flautas de chinos}, focuses on the ability of the jet-drive model of flute-like instruments to generate sounds of the same kind as \textit{sonido rajado} produced by these instruments.

As a first conclusion, a study of the sound produced by a \textit{flauta de chinos} has highlighted the quasiperiodic nature of \textit{sonido rajado}.

Secondly, time domain simulations show that not only the jet-drive model can produce, as transverse flutes (for example), periodic solutions associated to the first or second register of the instrument, but also quasiperiodic regimes, which present similarities with the \textit{sonido rajado} of prehispanic chilean flutes.

Moreover, an analysis of the system dynamics through its bifurcation diagram highlights the production mechanism of those quasiperiodic regimes.
Comparison between the sound produced by \textit{flautas de chinos} and the model dynamics thus suggests that the desired behaviour of such instruments would correspond to a quasiperiodic regime emerging from a direct Neimark-Sacker bifurcation of the periodic solution branch related to the first register of the instrument.

However, a quantitative comparison between experimental and synthesis sounds is impossible here as we study on the one hand the sound radiated by the real instrument whereas we consider, in the model, the internal acoustic field. But more importantly, both the fact that the precise values of different parameters of the model are unknown and the fact that different phenomena are not taken into account (for example, the jet is certainly turbulent \cite{Blanc_10}) limit such a quantitative comparison. These elements probably explain why significant differences remain between the real sound and the synthesis one, such as, for quasiperiodic regimes, the variation of the modulation frequency with the jet velocity.

\bibliography{smac_QP_biblio}

\begin{thebibliography}{10}
\providecommand{\url}[1]{#1}
\csname url@samestyle\endcsname
\providecommand{\newblock}{\relax}
\providecommand{\bibinfo}[2]{#2}
\providecommand{\BIBentrySTDinterwordspacing}{\spaceskip=0pt\relax}
\providecommand{\BIBentryALTinterwordstretchfactor}{4}
\providecommand{\BIBentryALTinterwordspacing}{\spaceskip=\fontdimen2\font plus
\BIBentryALTinterwordstretchfactor\fontdimen3\font minus
  \fontdimen4\font\relax}
\providecommand{\BIBforeignlanguage}[2]{{%
\expandafter\ifx\csname l@#1\endcsname\relax
\typeout{** WARNING: IEEEtran.bst: No hyphenation pattern has been}%
\typeout{** loaded for the language `#1'. Using the pattern for}%
\typeout{** the default language instead.}%
\else
\language=\csname l@#1\endcsname
\fi
#2}}
\providecommand{\BIBdecl}{\relax}
\BIBdecl

\bibitem{Wright}
H.~Wright and D.~Campbell, ``Analysis of the sound of chilean pifilca flutes,''
  \emph{The Galpin Society Journal}, vol.~51, pp. 51--63, 1998.

\bibitem{Blanc_10}
F.~Blanc, P.~de~la Cuadra, B.~Fabre, G.~Castillo, and C.~Vergez, ``Acoustics of
  the "flautas de chinos",'' in \emph{Proc. of 20th Int. Symposium on Music
  Acousitcs}, Sydney, 2010.

\bibitem{Coltman_06}
J.~Coltman, ``Jet offset, harmonic content, and warble in the flute,''
  \emph{Journal of the Acoustical Society of America}, vol. 120, no.~4, pp.
  2312--2319, 2006.

\bibitem{Fletcher_76}
N.~Fletcher, ``Sound production by organ flue pipes,'' \emph{Journal of the
  Acoustical Society of America}, vol.~60, no.~4, pp. 926--936, 1976.

\bibitem{Nayfeh}
A.~Nayfeh and B.~Balachandran, \emph{Applied Nonlinear Dynamics}.\hskip 1em
  plus 0.5em minus 0.4em\relax Wiley, 1995.

\bibitem{Gibiat}
V.~Gibiat, ``Phase space representations of acoustical musical signals,''
  \emph{Journal of Sound and Vibration}, vol. 123, no.~3, pp. 529 -- 536, 1988.

\bibitem{Zou}
Y.~Zou, ``Exploring reccurences in quasiperiodic dynamical systems,'' Ph.D.
  dissertation, Potsdam Univesity, 2007.

\bibitem{Fabre_Kergo}
A.~Chaigne and J.~Kergomard, \emph{Acoustique des instruments de musique
  (Acoustics of musical instruments)}.\hskip 1em plus 0.5em minus 0.4em\relax
  Belin (Echelles), 2008, ch.~10.

\bibitem{De_La_Cuadra_07}
P.~de~la Cuadra, C.~Vergez, and B.~Fabre, ``Visualization and analysis of jet
  oscillation under transverse acoustic perturbation,'' \emph{Journal of Flow
  Visualization and Image Processing}, vol.~14, no.~4, pp. 355--374, 2007.

\bibitem{Rayleigh}
J.~Rayleigh, \emph{The theory of sound second edition}.\hskip 1em plus 0.5em
  minus 0.4em\relax New York, Dover, 1894.

\bibitem{these_de_la_Cuadra}
P.~de~la Cuadra, ``The sound of oscillating air jets: Physics, modeling and
  simulation in flute-like instruments,'' Ph.D. dissertation, Stanford
  Univesity, 2005.

\bibitem{Nolle_98}
A.~Nolle, ``Sinuous instability of a planar jet: propagation parameters and
  acoustic excitation,'' \emph{Journal of the Acoustical Society of America},
  vol. 103, no.~6, pp. 3690--3705, 1998.

\bibitem{Coltman_76}
J.~Coltman, ``Jet drive mechanisms in edge tones and organ pipes,''
  \emph{Journal of the Acoustical Society of America}, vol.~60, no.~3, pp.
  725--733, 1976.

\bibitem{Verge_97}
M.~Verge, A.Hirschberg, and R.~Causs\'e, ``Sound production in recorder-like
  instruments. ii. a simulation model,'' \emph{Journal of the Acoustical
  Society of America}, vol. 101, no.~5, pp. 2925--2939, 1997.

\bibitem{Verge_94}
M.~Verge, R.~Causs\'e, B.~Fabre, A.Hirschberg, A.~Wijnands, and A.~van
  Steenbergen, ``Jet oscillations and jet drive in recorder-like instruments,''
  \emph{Acta Acustica united with Acustica}, vol.~2, pp. 403--419, 1994.

\bibitem{Segoufin_00}
C.~S\'egoufin, B.~Fabre, M.~Verge, A.~Hirschberg, and A.~Wijnands,
  ``Experimental study of the influence of the mouth geometry on sound
  production in a recorder-like instrument: windway length and chamfers,''
  \emph{Acustica united with Acustica}, vol.~86, pp. 649--661, 2000.

\bibitem{Fabre_96}
B.~Fabre, A.~Hirschberg, and A.~Wijnands, ``Vortex shedding in steady
  oscillation of a flue organ pipe,'' \emph{Acta Acustica united with
  Acustica}, vol.~82, no.~6, pp. 863--877, 1996.

\bibitem{Macaluso_11}
C.~Macaluso and J.~Dalmont, ``Trumpet with near-perfect harmonicity: Design and
  acoustic results,'' \emph{Journal of the Acoustical Society of America}, vol.
  129, no.~1, pp. 404--414, 2011.

\bibitem{CFA}
S.~Terrien, R.~Auvray, B.~Fabre, P.~Lagr\'ee, and C.~Vergez, ``Numerical
  resolution of a physical model of flute-like instruments: comparison between
  different approaches,'' in \emph{Proceedings of Acoustics 2012}, Nantes,
  France, 2012.

\bibitem{Krauskopf}
B.~Krauskopf, H.~Osinga, and J.~Galan-Vioque, \emph{Numerical continuation
  methods for dynamical systems}.\hskip 1em plus 0.5em minus 0.4em\relax
  Springer, 2007.

\bibitem{Doedel_81}
E.~J. Doedel, ``{AUTO: A} program for automatic bifurcation analysis of
  autonomous systems,'' \emph{Congressus Numerantium}, vol.~30, pp. 265--284,
  1981.

\bibitem{Cochelin_09}
B.~Cochelin and C.~Vergez, ``A high order purely frequency-based harmonic
  balance formulation for continuation of periodic solutions,'' \emph{Journal
  of Sound and Vibration}, vol. 324, no. 1-2, pp. 243--262, 2009.

\bibitem{manuel_Biftool}
K.~Engelborghs, ``{DDE B}iftool: a {M}atlab package for bifurcation analysis of
  delay differential equations,'' Katholieke Universiteit Leuven, Tech. Rep.,
  2000.

\bibitem{Barton_06}
D.~Barton, B.~Krauskopf, and R.~Wilson, ``Collocation schemes for periodic
  solutions of neutral delay differential equations,'' \emph{Journal of
  Difference Equations and Applications}, vol.~12, no.~11, pp. 1087--1101,
  2006.

\bibitem{Karkar_11}
S.~Karkar, C.~Vergez, and B.~Cochelin, ``Oscillation threshold of a clarinet
  model: a numerical continuation approach,'' \emph{Journal of the Acoustical
  Society of America}, vol. 131, no.~1, pp. 698--707, 2012.

\bibitem{bergeot2012prediction}
B.~Bergeot, A.~Almeida, C.~Vergez, and B.~Gazengel, ``Prediction of the dynamic
  oscillation threshold in a clarinet model with a linearly increasing blowing
  pressure,'' \emph{Arxiv preprint arXiv:1207.4636}, 2012.

\end{thebibliography}

\end{document}